\documentclass{nature}

\usepackage{graphicx}
\usepackage{amsmath}
\usepackage{amssymb}
\usepackage[colorlinks, citecolor=purple]{hyperref}
\usepackage{xcolor}
\usepackage{bm}
\usepackage{bbm}
\usepackage{stmaryrd}
\usepackage{caption}

\makeatletter
\let\saved@includegraphics\includegraphics
\AtBeginDocument{\let\includegraphics\saved@includegraphics}
\renewenvironment{figure}{\@float{figure}}{\end@float}
\makeatother

\newcommand{\ii}{{\mathrm{i}}}

\newcommand{\ee}{{\mathrm{e}}}

\usepackage[textwidth=180mm]{geometry}

\usepackage{lineno}

\begin{document}
\title{Computed tomography of propagating microwave photons}
\author{Qi-Ming Chen$^{1,\ast}$, 
Aarne Ker{\"a}nen$^{1}$, 
Aashish Sah$^{1}$, 
Mikko M{\"o}tt{\"o}nen$^{1,2,\ast}$}

\maketitle

\begin{affiliations}
\item QCD Labs, QTF Centre of Excellence, Department of Applied Physics, Aalto University, FI-00076 Aalto, Finland.
\item VTT Technical Research Centre of Finland Ltd. $\&$ QTF Centre of Excellence, P.O. Box 1000, 02044 VTT, Finland.

$^{\ast}$Corresponding authors. E-mails: qiming.chen@aalto.fi (Q.C.); mikko.mottonen@aalto.fi (M.M.)
\end{affiliations}

\begin{abstract}
Propagating photons serve as essential links for distributing quantum information and entanglement across distant nodes. 
Knowledge of their Wigner functions not only enables their deployment as active information carriers but also provides error diagnostics when photons passively leak from a quantum processing unit.
While well-established for standing waves, characterizing propagating microwave photons requires post-processing of room-temperature signals with excessive amplification noise.
Here, we demonstrate amplification-free Wigner function tomography of propagating microwave photons using a superconductor--normal-metal--superconductor bolometer based on the resistive heating effect of absorbed radiation.
By introducing two-field interference in power detection, the bolometer acts as a sensitive and broadband quadrature detector that samples the input field at selected angles at millikelvin with no added noise.
Adapting the principles of computed tomography (CT) in medical imaging, we implement Wigner function CT by combining quadrature histograms across different projection angles and demonstrate it for Gaussian states at the single-photon level. 
Compressed sensing and neural networks further reduce the projections to three without compromising the reconstruction quality.
These results address the long-standing challenge of characterizing propagating microwave photons in a superconducting quantum network and establish a new avenue for real-time quantum error diagnostics and correction.
\end{abstract}

The Heisenberg uncertainty principle is a fundamental concept in quantum mechanics that dictates a quantum system cannot simultaneously hold a definite position, $x$, and momentum, $p$. In quantum electrodynamics, the Wigner quasi-distribution function, $W(x,p)$, provides a useful tool for describing the uncertainty principle, which facilitates measurement of the field operator in Weyl's symmetric order. Specific to the microwave regime, Wigner function tomography of cavity-confined photons has been realized by combining displacement operations and parity measurements using an ancilla qubit with dedicated control and readout circuitry \cite{Hofheinz2009, Sun2014, Kirchmair2013}. However, it is more challenging to characterize propagating microwave photons with only few successful demonstrations in the literature \cite{Menzel2010, Eichler2011, Eichler2011b, Menzel2012, Zhong2013, Flurin2015, Fedorov2016, Chen2023}.

At the single-microwave-photon level, the overwhelmingly large amplification noise obscures the actual signal of interest and forces photon statistics to be inferred mathematically after noise deconvolution \cite{Eichler2012}. A successful implementation of photon tomography relies on the precise characterization of the amplification gain and the added noise. Recent advances in superconducting quantum circuits enable photon number counting of pulsed propagating microwaves at millikelvin temperatures and thus circumvent the amplification noise. The propagating photons at a prescribed frequency of interest may be either resonantly absorbed \cite{Chen2011, Inomata2016, Opremcak2018, Lescanne2020} or dispersively reflected \cite{Kono2018, Besse2018, Wang2022} by a precisely designed cavity, followed by a sequence of parity measurements with a high-coherence ancilla qubit. Combining photon parity measurements with displacement operations in each duty cycle, this approach implements an active detector for Wigner function tomography without added noise \cite{Dassonneville2020, Besse2020}. Yet, the qubit-based detectors typically have a narrow measurement bandwidth and require a considerable number of parity measurements and dead time between them. Efficient characterization of propagating microwave photons with a passive and broadband cryogenic detector remains an unmet need for the development of future quantum networks and fault-tolerant error correction protocols in the microwave regime.

\begin{figure}
  \centering
  \includegraphics[width=18cm]{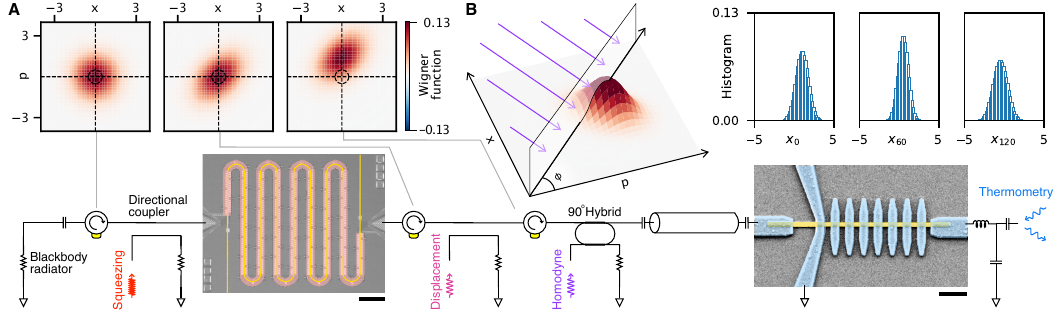}
  \linespread{1.1}
  \caption{{\bf Experiment overview.} 
{\bf A} Step-by-step generation of a broadband Gaussian state. A continuous beam of thermal photons is first generated from a blackbody radiator (left), then squeezed by a Josephson traveling-wave parametric amplifier (TWPA, middle), and finally displaced by a directional coupler (right) before measurement. 
Shown are Wigner functions at different stages, where $x$ and $p$ are the position and momentum variables, respectively.
{\bf B} Protocol for Wigner function computed tomography (CT). By controlling the phase of the homodyne field, we project the Wigner function into quadrature histograms at different angles $\phi$. 
The quadrature mean and variance, $\langle \hat{X}_{\phi} \rangle$ and $\langle (\Delta \hat{X}_{\phi})^2\rangle$, are simultaneously measured with a superconductor-normal metal-superconductor (SNS) bolometer, by sweeping a weak thermometry field at sub-gigahertz frequencies.
At the bottom, we show the simplified experimental setup with embedded micrographs of a reference TWPA (left) and SNS bolometer (right). 
The scale bar denotes $2\,{\rm mm}$ for the TWPA, with the junction array highlighted in red and flux bias in yellow. 
The scale bar denotes $2\,{\rm \mu m}$ for the bolometer, with the materials Al and Au$_{x}$Pd$_{y}$ highlighted in blue and yellow, respectively.
\vspace{1cm}
}
\end{figure}

Though resistive elements are generally avoided in superconducting quantum circuits, resistors allow for an exceptionally wide measurement bandwidth that is ideal for photon detection \cite{Wei2008, Govenius2014, Karimi2020, Lee2020}. Utilizing the resistive heating effect of the absorbed radiation \cite{Govenius2016, Kokkoniemi2019, Kokkoniemi2020, Gunyho2024}, recent work demonstrates that a superconductor--normal-metal--superconductor (SNS) bolometer with integrated radio-frequency readout can simultaneously measure the mean photon number and variance of microwave radiation \cite{Keranen2025}. The SNS bolometer is a passive and broadband detector that integrates the input power continuously at millikelvin without added noise. Here, we introduce two-field interference in bolometry, transforming this power detector into a noiseless quadrature measurement apparatus. By sweeping the phase difference between these two fields, we project the two-dimensional Wigner function into a set of marginal distributions taken at defined angles \cite{Lvovsky2009, Mallet2011}. The Wigner function is reconstructed following the well-established methodology of computed tomography (CT) in medical imaging. Thus, we achieve complete characterization of propagating microwave photons with an SNS bolometer, with the Wigner function CT framework further allowing a multitude of image processing methods that reduce the measurement overhead and enhance the reconstruction quality \cite{Zeng2023}. 

In our experiment, we use a $90^{\circ}$ hybrid as the microwave beam splitter that combines the bosonic input field, $\hat{a}$, with a large coherent homodyne field, $|\beta\rangle$ with $\beta=|\beta|\ee^{\ii \phi}$, before bolometry. The photon number of the combined field, $\hat{c}$, includes that of the transmitted input and reflected homodyne fields and, most importantly, an interference term between them. For a sufficiently large $|\beta|^{2}$, photon number measurement of the combined field, $\hat{c}$, is equivalent to a quadrature measurement of the input field, $\hat{a}$. On average, we have
\begin{eqnarray}
    \langle \hat{n}_{c} \rangle &=& \Gamma \langle \hat{n}_{a} \rangle
    + (1-\Gamma)|\beta|^2 + \sqrt{2\Gamma(1-\Gamma)|\beta|^2}\langle \hat{X}_{\phi+90} \rangle, 
    \label{eq:n}\\
    \langle (\Delta \hat{n}_{c})^2 \rangle &\approx & 
    (1-\Gamma)|\beta|^2 
    \big[ (1-\Gamma)
    + 2\Gamma\langle (\Delta \hat{X}_{\phi+90})^2 \rangle \big], 
    \label{eq:dn}
\end{eqnarray}
where $\Gamma$ is the transmissivity of the beam splitter, and $\hat{X}_{\phi} = (\hat{a}^{\dagger}e^{+\ii\phi} + \hat{a}e^{-\ii\phi})/\sqrt{2}$ is the quadrature operator. When choosing $\phi=0^{\circ}$ and $90^{\circ}$, we obtain the position and momentum operators, $\hat{X}_{0}\equiv \hat{X}$ and $\hat{X}_{90}\equiv \hat{P}$, respectively, with the canonical commutation relation: $\llbracket \hat{X}, \hat{P} \rrbracket = \ii$ with $\hbar=1$. In addition, $\langle \hat{n}_{y} \rangle$ and $\langle (\Delta \hat{n}_{y})^2 \rangle$ are the mean photon number and variance of the field $\hat{y}$. Note that the photons in a continuous beam of radiation have units ${\rm \left(s\times Hz\right)^{-1}}$, which will not be explicitly emphasized in the following discussions \cite{Blow1990}. Assuming a Gaussian state of the input field, as it is the common choice of continuous-variable states in a quantum network, the knowledge of $\langle \hat{X}_{\phi} \rangle$ and $\langle (\Delta \hat{X}_{\phi})^2 \rangle$ fully defines the marginal distribution of the Wigner function at angle $\phi$, i.e., $h_{\phi}\left(x_{\phi}\right) = \int {\rm d}p_{\phi}\, W\left(x_{\phi}\cos\phi - p_{\phi}\sin\phi, x_{\phi}\sin\phi + p_{\phi}\cos\phi \right)$, where $x_{\phi}$ and $p_{\phi}$ correspond to the eigenvalues of $\hat{X}_{\phi}$ and $\hat{X}_{\phi+90}$, respectively. This projection process can be precisely described by the Radon transform in medical imaging \cite{Zeng2023}. It allows the reconstruction of the Wigner function by sweeping $\phi$ at a certain interval following the well-established methodology of CT, as illustrated in Fig.\,1B. 

The Gaussian states studied in this experiment are defined as squeezed and displaced thermal states of the form $\hat\rho=\hat D(\alpha)\hat S(\zeta)\hat\rho_{T}\hat S^{\dagger}(\zeta)\hat D^{\dagger}(\alpha)$, where $\hat\rho_{T}$ is a thermal state with mean photon number $\bar{n}_{T}$, and $\hat S(\zeta)$ and $\hat D(\alpha)$ are the squeezing and displacement operators, respectively. In our experiment, we generate arbitrary Gaussian states step-by-step by combing a blackbody radiator, a Josephson traveling-wave parametric amplifier (TWPA), and a directional coupler, with the simplified setup shown in Fig.\,1A. The blackbody radiator is a primary source of thermal photons, with $\bar{n}_{T}$ controlled by varying its local temperature \cite{Mariantoni2010}. The squeezing and displacement parameters, $\zeta$ and $\alpha$, are controlled by applying a double-frequency squeezing field to the TWPA \cite{Perelshtein2022, Esposito2022, Qiu2023} and a single-frequency displacement field to the directional coupler \cite{Fedorov2016}, respectively. After interfering the input with the coherent homodyne fields, we measure the mean photon number, $\langle \hat{n}_{c} \rangle$, and variance, $\langle (\Delta \hat{n}_{c})^2 \rangle$, using an SNS bolometer \cite{Keranen2025}. The information is encoded in the resonance frequency, $\mu$, and Gaussian broadening, $\sigma^2$, of the averaged complex reflection spectrum of the thermometer, which is read out by sweeping a weak thermometry field at sub-gigahertz frequencies \cite{Chen2023b} (see Methods). In our experiment, the measurement bandwidth of the bolometer is constrained by the on-chip filter in front of the absorber, which is centered at $f_{0}=8.43\,{\rm GHz}$ with a linewidth of ${\rm FWHM}=133\,{\rm MHz}$, see Fig.\,1B.

\begin{figure}
  \centering
  \includegraphics[width=18cm]{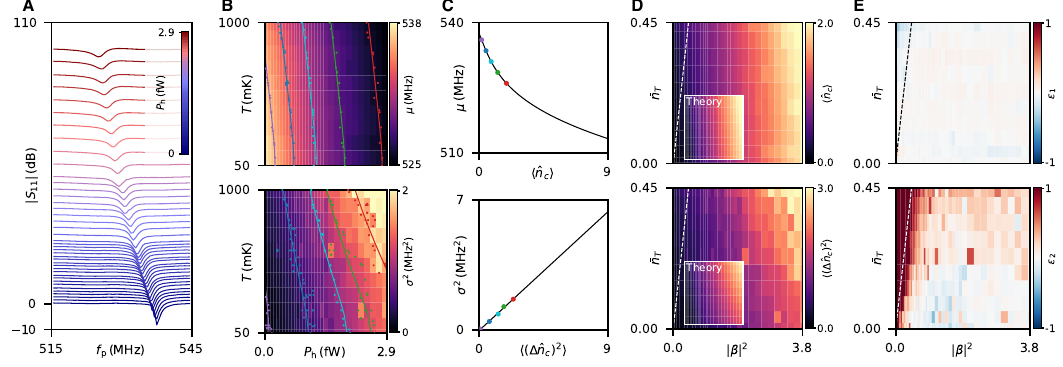}
  \linespread{1.1}
  \caption{
{\bf Bolometry of propagating thermal photons.} 
{\bf A} Reflection amplitude of the thermometer signal, $|S_{11}|$, as a function of its frequency for different indicated homodyne powers, $P_{\rm h}$. 
The thermal input field is prepared at a radiation temperature of $T=50\,{\rm mK}$. 
Spectra at different powers are vertically offset, and the backgrounds are subtracted using a polynomial fit.
{\bf B} The resonance frequency, $\mu$, and the Gaussian broadening, $\sigma^2$, of the thermometry spectra versus $P_{\rm h}$ and $T$.
Dots highlight data along five evenly spaced contours, and curves are the corresponding polynomial fits.
{\bf C} Measured $\mu$ as a function of the mean photon number $\langle \hat{n}_{c} \rangle$ and $\sigma^2$ as a function of the photon-number variance $\langle (\Delta \hat{n}_{c})^2 \rangle$ at the contours of panel B. The solid curves represent third-order and first-order polynomial fits.
{\bf D} Experimentally extracted $\langle \hat{n}_{c} \rangle$ and $\langle (\Delta \hat{n}_{c})^2 \rangle$ using the curves of panel C as a calibration versus the mean input photon number, $\bar{n}_{T}$, and homodyne photon number, $|\beta|^{2}$. 
Insets show theoretical expectations over identical parameter ranges and color maps.
{\bf E} Deviations of the measured quadrature mean, $\varepsilon_1 = \langle \hat{X}_{90} \rangle$, and variance, $\varepsilon_2 = \langle (\Delta \hat{X}_{90})^2 \rangle - (2\bar{n}_{T}+1)/2$, as compared to theory.
Dashed curves in D and E indicate $|\beta|^2 = \bar{n}_{T}$. All axes and color bars are shown in linear scale.
}
\end{figure}

\subsection{Bolometer characterization.}
First, we generate thermal states by heating the blackbody radiator from $T=50\,{\rm mK}$ to $1\,{\rm K}$ in $100\,{\rm mK}$ steps. At each temperature, we sweep the homodyne power from $P_{\rm h}=-138$ to $-112\,{\rm dBm}$ (estimated at the beam splitter input) and record the averaged reflection coefficient of the thermometer, as illustrated in Fig.\,2A. Depending on the specific combination of the thermal radiation and the homodyne field, the spectral lineshapes show different resonance frequencies, $\mu$, and Gaussian broadenings, $\sigma^{2}$, which are obtained by fitting to the Voigt profile. In Fig.\,2B, we highlight the contours of $\mu$ and $\sigma^2$, corresponding to unknown but constant values of the photon number $\langle \hat{n}_{c} \rangle$ and variance $\langle\left(\Delta \hat{n}_{c}\right)^2 \rangle$, respectively. By fitting these contours as a function of $T$ and $P_{\rm h}$, we obtain the best estimation of the transmissivity as $\Gamma=0.49$ ($0.1\,{\rm dB}$ imbalance) and an $\eta_1=-3.4\,{\rm dB}$ correction of the homodyne power. Here, we consider $\eta_{0}=-6.5\,{\rm dB}$ insertion loss from the radiator to the beam splitter input based on the datasheet values detailed in Supplementary Note\,3. We note that $\eta_{0}$ is only used for the characterization purpose, where a $\pm 0.5\,{\rm dB}$ difference does not significantly change the characterization result. In principle, this setup can be self-calibrated by using customized hybrids for cryogenic applications, where $\Gamma$ can be precisely engineered with $0.03\,{\rm dB}$ imbalance \cite{Mariantoni2010}. The only free parameter, $\eta_{1}$, can be adjusted by rescaling the reconstructed Wigner function with respect to the Heisenberg limit of a vacuum reference state. 

The described fitting procedure also determines $\langle \hat{n}_{c} \rangle$ and $\langle \left(\Delta \hat{n}_{c}\right)^2 \rangle$ at the selected contours, establishing calibration curves between $\mu$ and $\langle \hat{n}_{c} \rangle$, and $\sigma^2$ and $\langle \left(\Delta \hat{n}_{c}\right)^2 \rangle$. As shown in Fig.\,2C, we observe that $\langle \hat{n}_{c} \rangle$ follows a cubic dependence on $\mu$, while $\langle (\Delta \hat{n}_{c})^2 \rangle$ scales linearly with $\sigma^2$. The calibration curves can be verified by comparing the extracted values of $\langle \hat{n}_{c} \rangle$ and $\langle \left(\Delta \hat{n}_{c}\right)^2 \rangle$ with the theoretical expectations. Figure\,2D compares the consistency between them over the entire range of the thermal photon number $\bar{n}_{T}$ and the homodyne photon number $|\beta|^2$. Here, $\bar{n}_{T}=\eta_{0}/\left\{\exp\left[h f_{0}/(k_{\rm B}T)\right]-1\right\}$, where $h$ and $k_{\rm B}$ are Planck and Boltzmann constants, respectively, and $|\beta|^2 = \eta_{1}P_{\rm h}/({\rm FWHM}\times hf_{0})$. 

To determine a suitable homodyne photon number that validates the large-$|\beta|^2$ approximation in Eq.\,\eqref{eq:dn}, we compare the experimentally extracted $\langle \hat{X}_{90} \rangle$ and $\langle (\Delta \hat{X}_{90})^2 \rangle$ with the theoretical expectations for thermal states, i.e., $0$ and $(2\bar{n}_{T}+1)/2$, respectively. As shown in Fig.\,2E, the inaccuracy of the quadrature mean, $\langle \hat{X}_{90}\rangle$, remains below $0.1$ in the entire parameter range, whereas the extracted quadrature variance, $\langle (\Delta \hat{X}_{90})^2 \rangle$, exhibits a large discrepancy from theory when $|\beta|^{2} \leq \bar{n}_{T}$. With the increase of $|\beta|^{2}$, we observe a consistent decrease of the deviation that converges to zero at $|\beta|^{2} \rightarrow \infty$. We find that choosing $|\beta|^2 \gtrsim 10\bar{n}_{T}$ keeps the overall deviation below $0.2$ aside from several outliers at random places. 

\begin{figure}
  \centering
  \includegraphics[width=18cm]{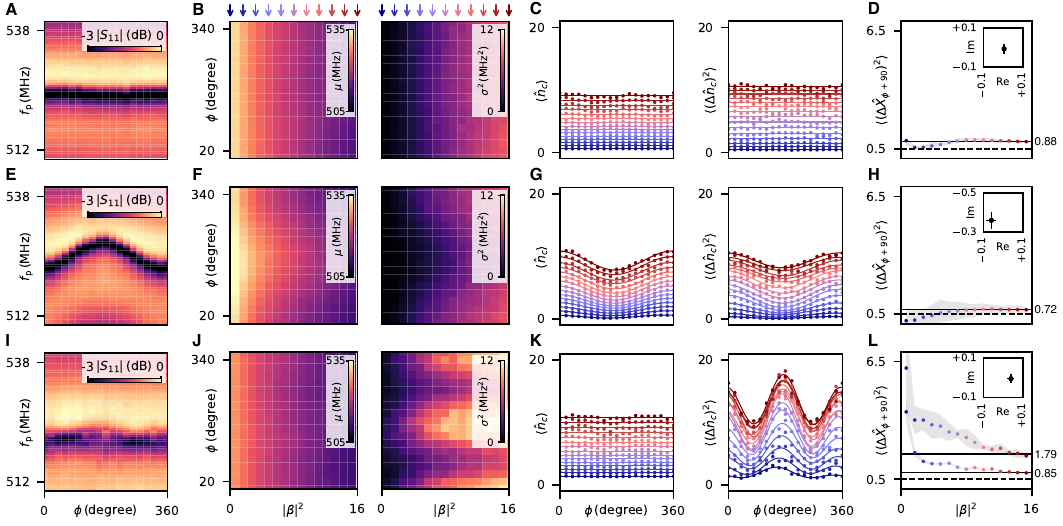}
  \linespread{1.1}
  \caption{{\bf Quadrature bolometry for symmetric Gaussian states.} 
{\bf A} Reflection amplitude of the thermometer, $|S_{11}|$, versus projection angle, $\phi$, for a thermal input state with $\bar{n}_{T}=0.45$, measured at a fixed homodyne photon number $|\beta|^2=3.8$.
The background is subtracted using a polynomial fit.
{\bf B} Resonance frequency, $\mu$, and the Gaussian broadening, $\sigma^2$, of the thermometry lineshape versus $\phi$ and $|\beta|^{2}$, obtained from Voigt-profile fit.
{\bf C} Mean photon number, $\langle \hat{n}_{c} \rangle$, and variance, $\langle (\Delta \hat{n}_{c})^2 \rangle$, of the input field as functions of $\phi$, evaluated at different $|\beta|^2$ as indicated in B. 
The solid curves show trigonometric fits.  
{\bf D} Mean value of the quadrature variance, $\langle (\Delta \hat{X}_{\phi+90})^2 \rangle$ over $\phi$ as a function of $|\beta|^2$, with the gray area bounded by the extrema. 
The solid and dashed lines indicate the mean value over the highest three $|\beta|^2$ and the Heisenberg limit of a vacuum state, respectively.
The inset shows the extracted displacement parameter $\alpha$, with the bars representing one standard deviation over $|\beta|^{2}$. 
The standard error of their mean is smaller than the marker size.
{\bf E}--{\bf H} As A--D but for a coherent input state without heating the blackbody radiator. From panel H, we obtain $\alpha = (-0.06 -\ii 0.36)$. 
The extracted $0.72$ quadrature variance is slightly larger than the Heisenberg limit ($0.5$), indicating a finite thermal population of $0.22$.
{\bf I}--{\bf L} As A--D but for a squeezed input state without heating the blackbody radiator. 
In L, we show two data points at each $|\beta|^2$, which represent the quadrature variances at squeezing angles ($\phi\in\{0^{\circ},\,180^{\circ}\}$) and anti-squeezing angles ($\phi\in\{90^{\circ},\,270^{\circ}\}$). 
The extracted values of $1.79$ and $0.85$ indicate a squeezing parameter of $\zeta=0.19$ ($1.6\,{\rm dB}$ squeezing).
All axes and color bars are shown in linear scale.
}
\end{figure}

\subsection{Quadrature bolometry.}
Next, we sweep the phase of the homodyne field from $\phi=0^{\circ}$ to $360^{\circ}$ in $20^{\circ}$ steps to measure the marginal distributions of the Wigner function at different angles. Here, we actively compensate for the uneven insertion loss of the phase shifter by adjusting the homodyne photon number, $|\beta|^2$, accordingly. For the thermal state prepared at $T=1\,{\rm K}$, corresponding to a thermal population of $\bar{n}_{T}=0.45$ at the beam splitter input, we observe a phase invariant thermometry lineshape at each fixed $|\beta|^2$, as shown in Fig.\,3A. Fitting each lineshape with a Voigt profile, we obtain maps of the resonance frequency, $\mu$, and Gaussian broadening, $\sigma^2$, over the control parameters $\phi$ and $|\beta|^2$. As expected from the rotational symmetry of both $\langle \hat{X}_{\phi} \rangle$ and $\langle (\Delta \hat{X}_{\phi})^2 \rangle$, the observed values of $\mu$ and $\sigma^2$ and the extracted photon number, $\langle \hat{n}_{c} \rangle$, and variance, $\langle (\Delta \hat{n}_{c})^2 \rangle$ are also independent of $\phi$, as shown in Figs.\,3B and 3C, respectively. We then calculate the quadrature mean, $\langle \hat{X}_{\phi} \rangle$, and variance, $\langle (\Delta \hat{X}_{\phi})^2 \rangle$, at different $|\beta|^2$ by using Eqs.\,\eqref{eq:n} and \eqref{eq:dn}. From the quadrature mean, we extract the mean value of the displacement $\alpha =(-0.01 + \ii0.01)$ with root-mean-square deviations of $0.01$ and $0.02$ in real and imaginary parts, respectively, which agrees with the expected vanishing mean for a thermal state. The quadrature variance converges to $0.88$ with increasing $|\beta|^2$, which is slightly below the theoretical value ($0.95$) due to the imperfection of bolometer characterization, see Fig.\,3D. 

We repeat the measurement procedure for a coherent input field that is generated by applying an $8.43\,{\rm GHz}$ displacement field at $-109\,{\rm dBm}$ (estimated at the directional coupler input). This state breaks the rotational symmetry of $\langle \hat{X}_{\phi} \rangle$ but preserves that for $\langle (\Delta \hat{X}_{\phi})^2 \rangle$. Correspondingly, we observe a $360^{\circ}$-periodic shift of the thermometry lineshape with its linewidth almost independent of $\phi$, see Fig.\,3E. The Voigt-profile fits confirm this observation with an oscillatory $\mu$, and correspondingly an oscillatory $\langle \hat{n}_{c} \rangle$, as shown in Figs.\,3F and 3G, respectively. These results indicate a mean value of the displacement $\alpha=(-0.06 - \ii 0.36)$ with root-mean-square deviations of $0.03$ and $0.04$ for the real and imaginary parts, respectively. However, we also observe $360^{\circ}$-periodic oscillations of $\sigma^2$ and $\langle (\Delta \hat{n}_{c})^2 \rangle$, which are not captured by Eq.\,\eqref{eq:dn}. In fact, these oscillations are precisely described by the exact formula of $\langle (\Delta \hat{n}_{c})^2 \rangle$ without large-$|\beta|^{2}$ approximation (see Methods). As illustrated in Fig.\,3H, these oscillations contribute marginally to the evaluation of $\langle (\Delta \hat{X}_{\phi})^2 \rangle$ as $|\beta|^{2}$ increases. The quadrature variance converges to $0.72$ at the maximum available $|\beta|^{2}$, which is slightly larger than the Heisenberg limit of coherent states ($0.5$). This result indicates a finite thermal population of $\bar{n}_{T}=0.22$ in the input field. We attribute these thermal photons to the blackbody radiation of the attenuation stages in the displacement line, which degrades the purity of the coherent state. 

As our last example of quadrature bolometry, we measure a squeezed input state that is generated by applying a $16.86\,{\rm GHz}$ squeezing field at $-103\,{\rm dBm}$ (estimated at the TWPA input). This field results in a squeezing operator on the vacuum field at $8.43\,{\rm GHz}$. However, due to the broadband nature of TWPA, the off-resonant vacuum photons are parametrically amplified simultaneously, leading to an overall increase of the thermal population and decrease of the squeezing level at the beam splitter input. The resulting state preserves the rotational symmetry of $\langle \hat{X}_{\phi} \rangle$ but breaks that of $\langle (\Delta \hat{X}_{\phi})^2 \rangle$, from which we expect a flat frequency shift over $\phi$ but an oscillatory linewidth with a $180^{\circ}$ periodicity. This is consistent with our observation in Fig.\,3I, together with the Voigt-profile fits of $\mu$ and $\sigma^2$ in Fig.\,3J, and the extracted $\langle \hat{n}_{c} \rangle$ and $\langle (\Delta \hat{n}_{c})^2 \rangle$ in Fig.\,3K. From the quadrature mean, we extract the mean value of displacement $\alpha=(0.04 - \ii0.00)$ with root-mean-square deviation of $0.02$ in  both real and imaginary parts. Because the quadrature variance is periodic with $\phi$, we illustrate their dependence on $|\beta|^2$ at the squeezing angles ($\phi \in \{0^{\circ},\,180^{\circ}\}$) and anti-squeezing angles ($\phi \in \{90^{\circ},\,270^{\circ}\}$), as shown in Fig.\,3L. Averaging over the three highest homodyne photon numbers, we obtain the maximum and minimum values of $\langle (\Delta \hat{X}_{\phi})^2 \rangle$ as $1.79$ and $0.85$, respectively, indicating a $1.6\,{\rm dB}$ squeezing along the $p$ quadrature. Their product implies a residual thermal population of $\bar{n}_{T}=0.73$, which is a trade-off inherent to broadband squeezing. 

\begin{figure}
  \centering
  \includegraphics[width=18cm]{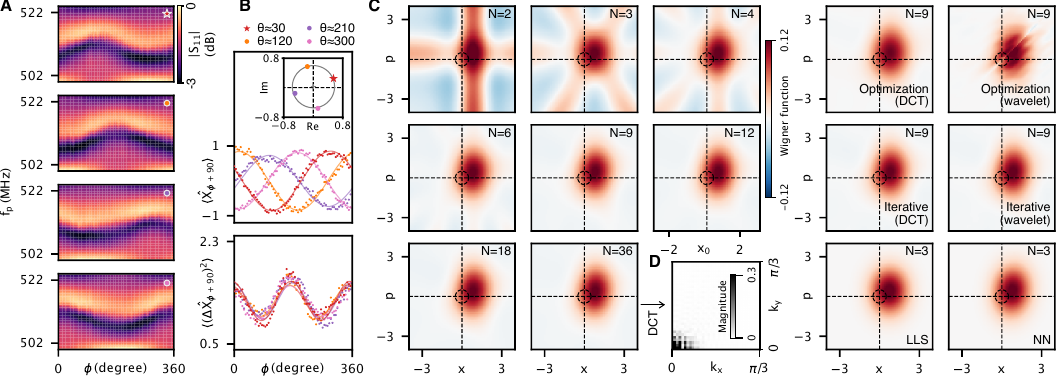}
  \linespread{1.1}
  \caption{{\bf Wigner function computed tomography (CT).} 
{\bf A} Reflection amplitude of the thermometer, $|S_{11}|$, versus the projection angle, $\phi$, for four Gaussian states. 
From top to bottom, the displacement parameter $\alpha$ rotates linearly with steps of $90^{\circ}$.
Backgrounds are subtracted using a polynomial fit.
{\bf B} Quadrature mean, $\langle \hat{X}_{\phi+90} \rangle$, and variance, $\langle (\Delta \hat{X}_{\phi+90})^2 \rangle$, as functions of $\phi$ for the Gaussian states in A.
The solid curves represent trigonometric fits with periods of $360^{\circ}$ and $180^{\circ}$, respectively.
Inset shows the extracted values of the displacement, $\alpha$. 
{\bf C} Wigner functions reconstructed from the first spectrum in A (marked by star), using Hilbert transform with no fitting parameter.
The projection angles are evenly chosen from $0^{\circ}$ to $180^{\circ}$ with $N$ steps. 
{\bf D} Discrete cosine transform (DCT) of the reconstructed Wigner function for $N=36$ versus the wave numbers $k_x$ and $k_y$. 
Shown are only the low-frequency components that reveal the sparsity of the Wigner function in the reciprocal space, while the high-frequency components are negligibly small.
{\bf E} Wigner functions obtained with sparse sampling ($N=9$), using compressed sensing with DCT and Daubechies wavelet bases. The reconstruction is carried out by either minimizing the $\ell^{1}$-norm of the sparsely represented Wigner function or by iteratively thresholding its significant components. 
{\bf F} Wigner functions obtained with ultra-sparse sampling ($N=3$), using linear least-squares (LLS) regression and neural network (NN). All axes and color bars are shown in linear scale.
}
\end{figure}

\subsection{Wigner function CT.}
Going beyond quadrature measurement with the SNS bolometer, we apply Wigner function CT to a general Gaussian state that is characterized by finite values of $\bar{n}_{T}$, $\zeta$, and $\alpha$. Such a state breaks the rotational symmetry of both $\langle \hat{X}_{\phi} \rangle$ and $\langle (\Delta \hat{X}_{\phi})^2 \rangle$, resulting in a thermometry spectrum with phase-dependent resonance frequency, $\mu$, and Gaussian broadening, $\sigma^2$. Because the TWPA inherently results in a finite $\bar{n}_{T}$, here we keep the blackbody radiator thermalized at the mixing chamber temperature and apply a $-103\,{\rm dBm}$ squeezing field and a $-108\,{\rm dBm}$ displacement field to the generator. The homodyne photon number is set to the maximum available value, $|\beta|^2=15.3$, in the pursuit of the large-$|\beta|^{2}$ approximation, while the homodyne phase is swept from $\phi=0^{\circ}$ to $360^{\circ}$ with $N$ steps. 

Figure\,4A shows the spectra of four Gaussian states which are varied by approximately $90^{\circ}$ in the displacement parameter. As expected, the quadrature bolometry shows $360^{\circ}$-periodic oscillations of their quadrature mean, $\langle \hat{X}_{\phi}\rangle$, with an approximately $90^{\circ}$ phase shift in between, see Fig.\,4B. The extracted values of the quadrature variance, $\langle (\Delta \hat{X}_{\phi})^{2}\rangle$, have a $180^{\circ}$ periodicity, but with their phases well aligned with each other, indicating a fixed squeezing angle in our experiment.  Depending on the specific choice of $\arg(\alpha)$, we observe a small asymmetry in $\langle (\Delta \hat{X}_{\phi})^{2}\rangle$ with respect to $\phi$, which originates from the neglected small terms of Eq.\,\eqref{eq:dn}. Further increasing $|\beta|^2$ or fitting the data with a trigonometric function of periodicity $180^{\circ}$ may eliminate the influence of asymmetry in the extracted quadrature variance. 

For illustration, we perform Wigner function CT on the Gaussian state with $\arg(\alpha) \approx 30^{\circ}$, with only half of the measurement data ($\phi < 180^{\circ}$) being used owing to redundancy. We discretize the phase space into $M\times M$ pixels with $M=101$, and apply the Hilbert transform to reconstruct the Wigner function with no fitting parameter (see Supplementary Note\,4). As the number of projections increases from $N=2$ to $36$, we observe a monotonic improvement of the reconstruction quality, as shown in Fig.\,4C. At $N=36$, we extract a Gaussian state with $\bar{n}_{T}=0.71$, $\zeta=0.16-\ii0.13$, and $\alpha=0.55+\ii0.25$. The observed thermal population and $1.8\,{\rm dB}$ squeezing level are slightly larger but consistent with the values $0.73$ and $1.6\,{\rm dB}$ extracted from Figs.\,3I and 3L, where the same power of the squeezing field is used. 

Following Shannon's celebrated theorem, the quality of the reconstruction increases with the number of samples until saturation occurs at the Nyquist limit. Considering that each projective measurement samples approximately $M$ locations along the selected quadrature, one may expect to see the saturation of the reconstruction quality at $N \approx M$. However, the observed saturation at $N \approx 12$ occurs an order of magnitude earlier than this expectation. The observed reduction of required measurements may be understood in the sparsity of the Wigner function in the reciprocal space, where the Hilbert transform is effectively carried out, as shown in Fig.\,4D. Indeed, by treating the reconstruction result as a gray-scale image, it is known from image compression practices that one can accurately define it with only a few significant components in the reciprocal space.

\subsection{Compressed sampling.}
The sparsity in reciprocal space motivates the exploration of Wigner function CT with even fewer projections. Here, we fix $N=9$ and apply compressed sensing (CS) methods that search for the Wigner function with the maximum sparsity in a given basis \cite{Donoho2006, Candes2006b} (see Supplementary Note\,5). We consider two bases that are widely used for image compression: the discrete cosine transform (DCT) basis in the JPEG standard and the Daubechies wavelet basis in JPEG 2000. The reconstruction is implemented by either minimizing the $\ell^{1}$-norm or iteratively thresholding the significant components of the sparsely represented Wigner function. As shown in Fig.\,4E, CS yields consistent reconstruction results with the Hilbert transform approach (Fig.\,4C) but with fewer number of projections. Among the implementations of compressed Wigner function CT, the DCT basis results in fewer spurious features than the wavelet basis, and the iterative methods outperform the direct optimizations in computational efficiency. 

Note that we have not explicitly assumed any form of the Wigner function in the reconstruction process, but only assumed that the histograms taken at each $\phi$ have a Gaussian distribution. Assuming a Gaussian form of the Wigner function imposes constraints among different $\phi$, and indicates a minimum number of $N=3$ projective measurements for compressed Wigner function CT. Here, we explore the possibility of such ultra-sparse sampling protocol with two noise removal approaches (see Supplementary Note\,6). On the one hand, we apply the linear least-squares (LLS) regression to fit $\bar{n}_{T}$, $\xi$, and $\alpha$, and then compute the Wigner function from the Gaussian-state model. On the other hand, we train a three-layer neural network (NN) with $32768$ simulated Gaussian states, and then apply it to the experimental data for quantum state tomography. As shown in Fig.\,4F, these methods provide consistent results with each other and achieve a comparable reconstruction quality as the normal- and sparse-sampling approaches (Figs.\,4C and 4E). A possible advantage of the NN-approach is the model-free architecture that is suitable for future experiments with non-Gaussian states. In addition, training the NN directly with the experimental data would enable identification of the systematic noise and result in a more robust reconstruction. 

The demonstrated Wigner function CT combines the strengths of cryogenic thermal detectors and medical imaging methodologies in the platform of superconducting quantum circuits, offering a transformative approach to characterize propagating microwave photons with speed-up by classical computational efficiency. This speed-up is exemplified by our sparse and ultra-sparse sampling protocols, enabling rapid Wigner function reconstruction with a minimum of three projective measurements. The SNS bolometer enables full-duty-cycle and amplification-free measurement of propagating microwave photons with a broad bandwidth that is beyond the reach of qubit-based detectors. The compact design of the SNS bolometer and its compatibility with the superconducting quantum technology further facilitate array integration and multiplexed readout\,\cite{Singh2024}. By interfering input photons with a coherent homodyne field, we enable quadrature measurements with the bolometer and translate the well-established CT method to quantum state tomography. These results should stimulate further interest in quantum networks at the centimeter and millimeter wavelengths, thereby extending the established strengths of superconducting quantum circuits in quantum computation and quantum simulation to the realm of quantum communication. 

An appealing future direction of Wigner function CT lies in measuring non-Gaussian states or photons that are entangled in multiple modes. The negativity of their Wigner functions is a critical resource for quantum advantage in computing and methodology, and serves as a fundamental test of quantum mechanics. Here, the key challenge is to generate higher-order moments of the input field in the photon number and variance operators at the bolometer input. This may be achieved by engineering the homodyne field in non-classical states or deploying multiple beam splitters and bolometers analogous to higher-order correlation measurements in quantum optics \cite{Cheng2022}. Beyond the interest of propagating microwave photons, the Wigner function CT technique may be adapted for the cryogenic characterization of other types of particles via transduction, such as surface-acoustic-wave phonons \cite{Forsch2019}. In addition to the use cases in quantum science, we note that the homodyne field effectively boosts the input power to a detectable level but leaves the overall noise level unchanged. It offers a viable way, in parallel with the continuous efforts to improve the sensitivity of the detector itself \cite{Gunyho2024b}, to detect weak signals at the yocto- or zepto-joule scale --- the trail of individual single-photon emission processes in the ${\rm GHz}$--${\rm THz}$ range. 

\section*{Methods}
\subsection{Experimental setup.}
The SNS bolometer shown in Fig.\,1 is fabricated on a $675\,{\rm \mu m}$-thick Si substrate with an additional $300\,{\rm nm}$ SiO${}_2$ layer on top. It consists of a $1\,{\rm \mu m} \times 150\,{\rm nm} \times 30\,{\rm nm}$ absorber as part of an Au$_{x}$Pd$_{y}$ nanowire with $x/y\approx 3/2$. This nanowire extends further underneath seven perpendicularly oriented Al islands. Each island has a volume of $300\,{\rm nm} \times 2.4\,{\rm \mu m} \times 100\,{\rm nm}$. The center points of every two adjacent islands are separated by $600\,{\rm nm}$. The short Au$_{x}$Pd$_{y}$ segments between the Al islands, operating as proximity Josephson junctions, are embedded in a sub-gigahertz resonator for thermometry \cite{Chen2023b}. The thermometer and absorber junctions are electronically separated by a ground in between. In addition, we fabricated a $7.5\,{\rm mm}$-long coplanar waveguide filter with $200\,{\rm nm}$-thick Nb at the absorber input to select the measurement bandwidth of interest. Calibration of this filter shows a central frequency at $f_{0}=8.43\,{\rm GHz}$ with a linewidth of ${\rm FWHM}=133\,{\rm MHz}$ \cite{Keranen2025}. More details about the fabrication and the operation principle of this bolometer can be found in the literature \cite{Govenius2016, Kokkoniemi2019}.

The Gaussian-state generator consists of a blackbody radiator for generating thermal photons, a TWPA for squeezing, and two directional couplers for delivering the squeezing and displacement fields  (see Supplementary Note\,1). A subsequent $90^{\circ}$ hybrid interferes the Gaussian input field with a coherent homodyne field, and the SNS bolometer implements the photon detection. The averaged thermometry spectrum is measured by sweeping the frequency of a weak thermometry field from $500$ to $550\,{\rm MHz}$ while recording the reflection coefficient. A detailed cryogenic setup is shown in Extended Data Fig.\,1.

At room temperature, we generate the squeezing, displacement, and homodyne fields from a single signal source to maximize the phase coherence among them. On the other hand, the sub-gigahertz thermometry field is split into two paths: the first one is used for thermometry and the second serves as the reference. The reflected thermometry field is down-converted to an intermediate frequency (IF) of $62.5\,{\rm MHz}$ and sampled at $250\,{\rm MS/s}$. We obtain a $32\,{\rm \mu s}$-long trace of the IQ quadratures in each measurement repetition, in which each data point is extracted from one single IF period and digitally filtered by a $500\,{\rm kHz}$ low-pass filter. The number of repetitions is nominally set to $5\times 10^{3}$, over which the data are averaged before transferring to the host computer. A detailed room-temperature setup is shown in Extended Data Fig.\,2.

\subsection{Projective bolometry.}
We describe the beam splitter interaction between the input and the homodyne fields, $\hat{a}$ and $\hat{b}$, as $\hat{c} = \sqrt{\Gamma}\hat{a} + \ii\sqrt{1-\Gamma}\hat{b}$. For a coherent homodyne field, $|\beta\rangle$ with $\beta=|\beta|\ee^{\ii\phi}$, the mean photon number and variance of the output field, $\hat{c}$, are given by (see Supplementary Note\,2)
\begin{eqnarray*}
    \langle \hat{n}_{c} \rangle &=& \Gamma \langle \hat{n}_{a} \rangle
    + (1-\Gamma)|\beta|^2 
    + \sqrt{2\Gamma(1-\Gamma)}|\beta|\langle \hat{X}_{\phi+90} \rangle, 
    \label{eq:n_exact}\\
    \langle (\Delta \hat{n}_c)^2 \rangle 
	&=& (1-\Gamma) |\beta|^2 \left[(1-\Gamma)
	+ 2\Gamma \langle (\Delta \hat{X}_{\phi+90})^2 \rangle \right] 
    + \Gamma^2 \langle (\Delta \hat{n}_{a})^2 \rangle 
    + \Gamma(1-\Gamma)\langle \hat{n}_{a} \rangle \nonumber\\
    &+& \sqrt{\Gamma(1-\Gamma)} |\beta| \left[\ii\Gamma\left(
    2\langle \hat{a}^{\dagger}\hat{a}^{\dagger}\hat{a} \rangle e^{\ii\phi} 
    - 2\langle \hat{a}^{\dagger}\hat{a}\hat{a} \rangle \ee^{-\ii\phi} \right)
    + \sqrt{2}\left(1-2\Gamma\langle \hat{n}_{a} \rangle \right) \langle \hat{X}_{\phi+90} \rangle \right]. 
    \label{eq:dn_exact}
\end{eqnarray*}
Here, the quadrature mean and variance are defined as
\begin{eqnarray*}
    \langle \hat{X}_{\phi} \rangle &=& (\hat{a}^{\dagger}e^{\ii\phi} + \hat{a}e^{-\ii\phi})/\sqrt{2},\\
    \langle (\Delta \hat{X}_{\phi})^2 \rangle &=& \left[
    \langle (\Delta \hat{a}^{\dagger})^2 \rangle e^{\ii2\phi}
    + \langle \left(\Delta \hat{a}\right)^2 \rangle \ee^{-\ii2\phi}
    + 2 \left(\langle \hat{n}_{a} \rangle - \langle \hat{a}^{\dagger} \rangle\langle \hat{a} \rangle\right) -1 \right]/2.
\end{eqnarray*}

On the other hand, the SNS bolometer responds to the absorbed radiation as a change in the thermometry lineshape. The reflection coefficient can be written in a similar way to the Voigt profile in laser spectroscopy,
\begin{eqnarray*}
    S_{11}^{\rm ave} &=& 1 - \frac{\ee^{\ii\phi}\gamma_{\rm c}}{2\sqrt{2\pi} \sigma} 
    {\rm erfcx}\left[\frac{(\gamma/2) + \ii\Delta}{2\sqrt{2}\pi \sigma}\right],
\end{eqnarray*}
where $\gamma_{\rm c}$ and $\gamma$ are the external and the total energy decay rates, respectively; $\phi$ describes the asymmetry of the circuit, $\Delta=2\pi(\mu-f_{\rm p})$ is the detuning between the resonance frequency, $\mu$, and the probe frequency, $f_{\rm p}$, $\sigma^2$ is the Gaussian broadening relative to a Lorentzian lineshape, and ${\rm erfcx}(\cdot)$ is the scaled complementary error function. On average, the mean resonance frequency, $\mu$, is determined by the mean input photon number, $\langle \hat{n}_{c} \rangle$, and the Gaussian broadening, $\sigma^2$, by the photon number variance, $\langle \left( \Delta \hat{n}_{c} \right)^{2}\rangle$ \cite{Keranen2025}. 

To extract $\langle \hat{X}_{\phi} \rangle$ and $\langle (\Delta \hat{X}_{\phi})^2 \rangle$, we first note that one may fairly approximate $\langle (\Delta \hat{n}_c)^2 \rangle$ by the leading-order term of $|\beta|^2$ for large homodyne power, see Eq.\,\eqref{eq:dn} of the main text. Thus, $\langle (\Delta \hat{X}_{\phi})^2 \rangle$ can be directly obtained from $\langle (\Delta \hat{n}_c)^2 \rangle$ with the calibrated parameters $\Gamma$ and $|\beta|^2$. On the other hand, we subtract $\langle \hat{n}_c\rangle$ by its mean value over $\phi$ to extract $\langle \hat{X}_{\phi} \rangle$, utilizing the property that $\langle \hat{X}_{\phi} \rangle$ has a zero mean over the projection angles. In this way, no fitting is carried out for obtained $\langle \hat{X}_{\phi} \rangle$ and $\langle (\Delta \hat{X}_{\phi})^2 \rangle$ in our experiment. This method is suitable for Wigner function CT of Gaussian states, where the projected histogram is fully determined by the quadrature mean, $\langle \hat{X}_{\phi} \rangle$, and variance $\langle (\Delta \hat{X}_{\phi})^2 \rangle$. For non-Gaussian states, we note that the exact expressions of $\langle \hat{n}_c\rangle$ and $\langle (\Delta \hat{n}_c)^2 \rangle$ contain the information of the following eight moments: $\langle \hat{a} \rangle$, $\langle \hat{a}^{\dagger} \rangle$, $\langle \hat{a}\hat{a} \rangle$, $\langle \hat{a}^{\dagger}\hat{a} \rangle$, $\langle \hat{a}^{\dagger}\hat{a}^{\dagger} \rangle$, $\langle \hat{a}^{\dagger}\hat{a}^{\dagger}\hat{a} \rangle$, $\langle \hat{a}^{\dagger}\hat{a}\hat{a} \rangle$, and $\langle \hat{a}^{\dagger}\hat{a}^{\dagger}\hat{a}\hat{a} \rangle$. By fitting each individual moment, it is possible to reconstruct the density matrix of an arbitrary continuous-variable state with a maximum of $3$ photons or a pure state with up to $4$ photons.

\subsection{Computed tomography.}
We define $x_{\phi}$ and $p_{\phi}$ as eigenvalues of the quadrature operators $\hat{X}_{\phi}$ and $\hat{X}_{\phi+90}$, respectively, which follow the canonical commutation relation: $\left[ \hat{X}_{\phi}, \hat{X}_{\phi+90} \right] = \ii$ with $\hbar=1$. The marginal probability density of the quantum state along the quadrature $x_{\phi}$ can be obtained by integrating the Wigner function along its conjugate variable \cite{Lvovsky2009} 
\begin{eqnarray*}
	h_{\phi}\left(x_{\phi}\right) = \int {\rm d}p_{\phi}\, W\left(x_{\phi}\cos\phi - p_{\phi}\sin\phi, x_{\phi}\sin\phi + p_{\phi}\cos\phi \right).
\end{eqnarray*}
This is the one-dimensional histogram obtained by a projective measurement at angle $\phi-90$. One can prove that the Wigner function can be recovered from $h_{\phi}\left(x_{\phi}\right)$ according to the so-called Hilbert transform (see Supplementary Note\,4)
\begin{eqnarray*}
	W\left(x,p\right) 
	&=& \frac{1}{2\pi}\int_{0}^{\pi} {\rm d}\phi \left[\frac{1}{\pi r} * \frac{\partial h_{\phi}(r)}{\partial r}\right]{\Big{|}_{r=x\cos\phi + p\sin\phi}},
\end{eqnarray*}
where $*$ is the convolution operator. The Hilbert transform is effectively performed in the reciprocal space according to the convolution theorem.

By discretizing the phase space into $M\times M$ pixels, we may describe the projective measurement as a linear map between the Wigner function and the one-dimensional histogram, ${\bf H}_{\phi}=\mathbbm{A}_{\phi}{\bf W}$. Here, ${\bf H}_{\phi}$ is the histogram at angle $\phi$, $\mathbbm{A}_{\phi}$ is the measurement matrix of dimension $M\times M^{2}$, and ${\bf W}$ is the Wigner function written in vector form. This is an under-determined problem that can only be solved by stacking ${\bf H}_{\phi}$ and $\mathbbm{A}_{\phi}$ over $N \geq M$ projective measurements. However, the compressed sensing (CS) method provides a feasible way to bypass the sampling theorem if ${\bf W}$ can be sparsely represented in a certain basis \cite{Donoho2006, Candes2006b}, i.e., ${\bf W}=\mathbbm{B}{\bf W'}$ where $\mathbbm{B}$ is a sparse representing matrix and ${\bf W'}$ is a sparse vector. It effectively searches for solutions of the linear system that minimize the $\ell^{1}$-norm of ${\bf W'}$ (see Supplementary Note\,5).

\section*{Data availability}
The data that support the findings of this study are available at \url{https://doi.org/10.24433/CO.9909617.v1}.  

\section*{Code availability}
The code for analyzing the data are available at \url{https://doi.org/10.24433/CO.9909617.v1}. 

\section*{References}
\bibliographystyle{naturemag}
\bibliography{BoloCT_REF.bib} 

\section*{Acknowledgments}
\noindent 
We thank Joonas Govenius and Visa Vesterinen for providing their work on the SNS bolometer and TWPA used in the experiments, Gheorghe-Sorin Paraoanu and Jukka Pekola for sharing laboratory instruments, and Andr{\'a}s M{\'a}rton Gunyh{\'o} for discussions.
This work is supported by the Academy of Finland Centre of Excellence program (No.\,336810), European Research Council under Advanced Grant ConceptQ (No.\,101053801),  Business Finland Foundation through Quantum Technologies Industrial (QuTI) project (No.\,41419/31/2020), Technology Industries of Finland Centennial Foundation, Jane and Aatos Erkko Foundation through the Future Makers program and through the SystemQ grant, Finnish Foundation for Technology Promotion (No.\,8640), and Horizon Europe programme HORIZON-CL4-2022-QUANTUM-01-SGA via the project OpenSuperQPlus100 (No.\,101113946).

\section*{Author Contributions Statement}
Q.C. carried out experiments and analyzed data. 
Q.C. and A.K. programmed FPGA and developed measurement software.
A.K. contributed to measurements and data analysis.
Q.C. and M.M. wrote the manuscript with significant contributions from A.S. 
M.M. contributed to experiment design and supervised the project. 

\section*{Competing Interests Statement}
M.M. declares that he is a co-founder and shareholder of IQM\,Finland\,Oy and QMill Oy. M.M. is an inventor of patents FI122887B, US9255839B2, JP5973445B2, and EP2619813B1 titled ``Detector of single microwave photons propagating in a guide". Other authors declare no competing interests.

\section*{List of Supplementary Materials}
Supplementary Notes\,1 -- 6\\
Supplementary Fig.\,S1\\
References 46 -- 50

\newpage
\makeatletter
\setcounter{figure}{0}
\renewcommand{\figurename}{\textbf{Extended Data Fig.}}
\renewcommand{\thefigure}{\arabic{figure}}
\renewcommand{\fnum@figure}{\textbf{\figurename\,\thefigure}}
\makeatother

\begin{figure}
  \centering
  \includegraphics[width=18cm]{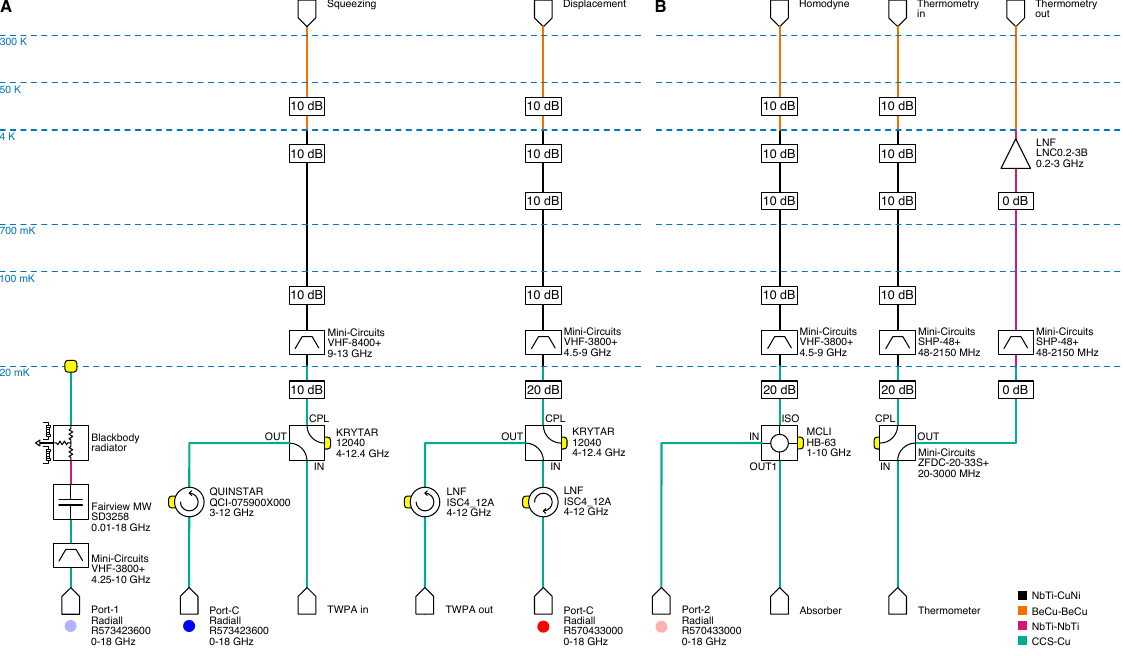}
  \linespread{1.1}
  \caption{{\bf Cryogenic setup.} 
{\bf A} Circuitry for Gaussian-state generator. Thermal photons are first generated by a blackbody radiator, then squeezed by a Josephson traveling-wave parametric amplifier (TWPA), and finally displaced by a directional coupler.
{\bf B} Circuitry for quadrature bolometry. The input field interferes with a homodyne field through a $90^{\circ}$ hybrid before reaching the absorber of the bolometer. The response is measured by sweeping a weak sub-gigahertz thermometry field that is applied to the thermometer. The input and output thermometry fields are separated by a directional coupler for reflection measurement.
}
\end{figure}

\begin{figure}
  \centering
  \includegraphics[width=18cm]{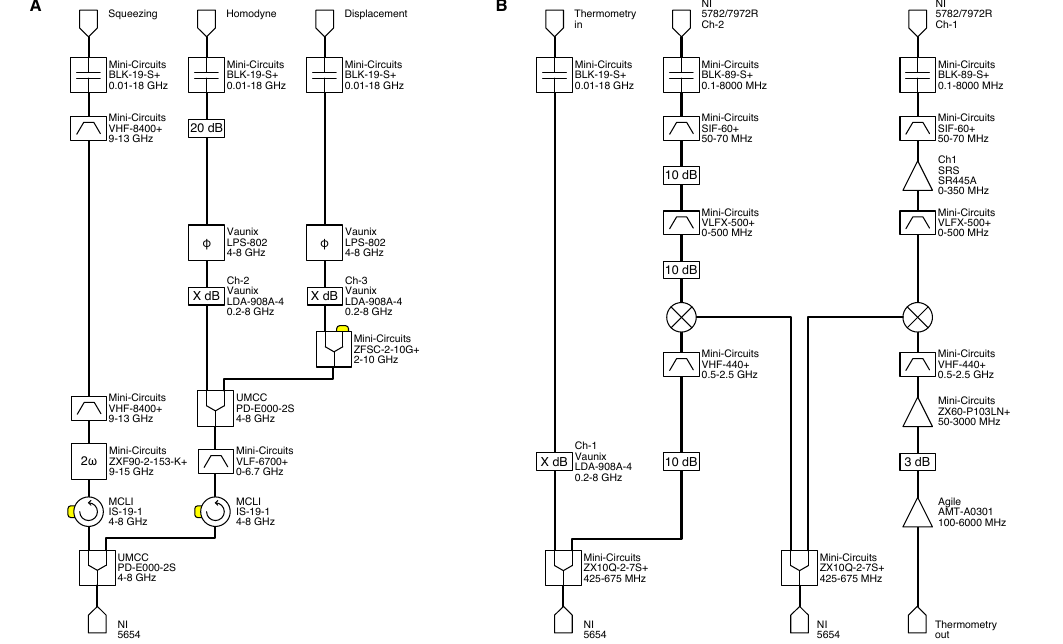}
  \linespread{1.1}
  \caption{{\bf Room-temperature setup.} 
{\bf A} Module for Gaussian-state generator. We generate the squeezing, displacement, and homodyne fields by splitting a single signal source to different paths to maximize the phase coherence among them. The relative phase and power are controlled by the two phase shifters and step attenuators.
{\bf B} Module for bolometry. The thermometry signal is split into two paths that are used for probing the resonance lineshape of the thermometer and providing the phase reference, respectively. The reflected thermometry field is amplified and down-converted for acquisition with the same local oscillator for the reference signal path. 
}
\end{figure}

\end{document}